**A tutorial for propensity score weighting for moderation analysis: An application examining smoking disparities among sexual minority adults**

Beth Ann Griffin,[a] Megan S. Schuler,[a] Matt Cefalu,[b] Lynsay Ayer [a], Mark Godley [c], Noah Griefer [d], Donna L. Coffman [e], Daniel McCaffrey [f]

[a] RAND Corporation, 1200 S Hayes St, Arlington, VA 22202

[b] RAND Corporation, 1776 Main St, Santa Monica, CA 90401

[c] Chestnut Health Systems, 448 Wylie Dr, Normal, IL 61761

[d] Institute for Quantitative Social Science, Cambridge, MA 02138

[e] University of South Carolina, Columbia, SC 29208

[f] ETS, 660 Rosedale Road, Princeton, NJ 08541 USA.

**Acknowledgments:** This research was financially supported through National Institutes of Health (NIH) grants (R01DA045049, PIs: Griffin/McCaffrey; P50DA046351, PI: Stein). NIH had no role in the design of the study, analysis, and interpretation of data nor in writing the manuscript.

**Word Count:** 4,859

**Number of Figures:** 2

**Number of Tables:** 3

**Number of Text Pages:** 13

**Conflicts of interest for all authors:**

Griffin – No conflicts

Schuler – No conflicts

Cefalu – No conflicts

Ayer - No conflicts

Godley - No conflicts

Griefer - No conflicts

Coffman - No conflicts

McCaffrey - No conflicts

**Abstract:**

*Objective*. To provide step-by-step guidance and STATA and R code for using propensity score (PS) weighting to estimate moderation effects.

*Research Design*. Tutorial illustrating the key steps for estimating and testing moderation using observational data. Steps include (1) examining covariate overlap across treatment groups within levels of the moderator, (2) estimating the PS weights, (3) evaluating whether PS weights improved covariate balance, (4) estimating moderated treatment effects, and (5) assessing sensitivity of findings to unobserved confounding. Our **illustrative** case study uses data from 41,832 adults from the 2019 National Survey on Drug Use and Health to examine if gender moderates the association between sexual minority status (e.g., lesbian, gay, or bisexual [LGB] identity) and adult smoking prevalence.

*Results*. **For our case study, there were no noted concerns about covariate overlap and we were able to successfully estimate the PS weights within each level of the moderator. Moreover,** balance criteria indicated that PS weights successfully achieved covariate balance for both moderator groups. PS weighted results indicated **there was significant evidence of moderation for the case study and sensitivity analyses demonstrated that results were highly robust for one level of the moderator but not the other.**

*Conclusions*. When conducting moderation analyses, covariate imbalances across levels of the moderator can cause biased estimates. As demonstrated in this tutorial, PS weighting within each level of the moderator can improve the estimated moderation effects by minimizing bias from imbalance within the moderator subgroups.

**Introduction**

Healthcare researchers are often interested in examining potential effect heterogeneity – i.e., moderation – when evaluating treatments, interventions, or health services received by individuals.[1,2] Understanding the influence that potential moderators may have on treatment effectiveness allows for greater understanding of the types of individuals who benefit the most (and least) from specific treatments and interventions, allowing more efficient targeting of resources.[3] In the context of observational data, moderation analyses – also referred to as treatment effect heterogeneity[4-6] and/or subgroup analyses[7,8] – allow researchers to explore important variation in the relationships between key exposures and health outcomes, leading to greater insights regarding population health and health disparities.

In terms of estimation, moderation is traditionally assessed in regression models by testing for an interaction effect between the potential moderator and the treatment/exposure or by using stratification methods. However, these standard approaches may yield biased estimates of the moderated effects if they do not adequately yield comparable treatment/exposure groups within each level of the moderator variable.[9-14] **While covariate adjustment can be useful to minize the potential bias that might arise from imbalance between the treatment and control group within each level of the moderator, relying on covariate adjustment alone can result in extrapolation of the outcome model and sensitivity to errors in the model specification.[15] It is more prudent to use causal inference approaches to ensure better comparability of the groups than covariate adjustment alone allows.[15-17]** Specifically, this tutorial focuses on moderation analyses using propensity score (PS) weighting to statistically balance treatment/exposure groups within all levels of the moderator with respect to potential confounders in order to mitigate confounding bias. To date, only one similar tutorial type paper has been published with attention to PS matching.[14] Similar guidance is needed for PS weighting.

Numerous PS weighting approaches for estimating moderated effects have been proposed; these approaches vary regarding the estimator of choice for the PS weights and/or the outcome model. Other methods in this space use more data-driven approaches (e.g., machine learning) to assess evidence of moderation in the presence of potential confounders.[18-32] **In fact, there has been an explosion of new methods in causal machine learning to estimate heterogenous treatment effects in recent years[21-33] that includes use of quasi-oracle, orthogonal statistical learning, generalized random forests, metalearners, and double (debiased) machine learning for estimation of treatment effect heterogeneity.** While numerous methods for moderation analyses have been proposed, there is no easy-to-use guide on how to implement these methods for health researchers that emphasizes key required assumptions and the importance of assessing the quality of the PS weights. The goal of this tutorial is to provide practitioners with step-by-step guidance and STATA and R code for conducting robust PS weighted analyses when estimating moderation effects. **Accordingly, we provide step-by-step guidance regarding PS weighting for moderation that includes (1) examining covariate overlap across treatment/exposure groups within levels of the moderator, (2) estimating the PS weights, (3) evaluating whether PS weights improved covariate balance within each moderator level, (4) estimating moderated treatment effects, and (5) assessing sensitivity of findings to potential unobserved confounding. This approach is broadly applicable to observational data – in particular, our case study illustrates the application of these methods to survey-weighted data.**

As an illustration of these methods, we present a case study that investigates whether gender moderates the association between sexual minority status (e.g., lesbian, gay, or bisexual (LGB) identity) and smoking. Numerous prior studies have demonstrated that sexual minority individuals report more frequent substance use, including smoking, compared to heterosexual individuals.[34-36] These disparities are often attributed to minority stress, namely the stigma, prejudice, and discrimination uniquely experienced by socially marginalized groups (e.g., LGB individuals).[37-40] Previous studies have highlighted gender differences in substance use,[41,42] including finding that disparities between sexual minority and heterosexual women are often larger in magnitude than those between sexual minority and heterosexual men.[34,43,44] In light of this, we frame gender as a potential moderator that may modify the strength of the relationship between sexual minority status and smoking in our case study.

**Methods**

**Data and measures**

*Study Population*

Data were from the 2019 National Survey on Drug Use and Health (NSDUH), an annual nationally-representative survey on drug use among the civilian, non-institutionalized US population ages 12 and older.[45] The sample size for the public-use NSDUH data is 56,136 for 2019 (65% response). Our sample is restricted to individuals ages 18 years and older who identified as heterosexual, lesbian/gay, or bisexual (n=41,832, including 3,618 LGB adults); respondents ages 12-17 are excluded as the NSDUH does not ask minors about sexual identity. Respondents who did not respond to the sexual identity question or answered, "don't know" (n=907, representing 2.1% of adult NSDUH respondents) are also excluded. This study was deemed exempt from review by RAND's institutional review board, as it involved de-identified survey data.

*Measures*

<u>Primary exposure, LGB status</u>: Sexual identity is assessed by asking: "Which one of the following do you consider yourself to be?" ("Heterosexual, that is, straight," "Lesbian or gay," "Bisexual," "Don't know"). An indicator for LGB status is defined as "1" for those who responded either "lesbian or gay" or "bisexual" and "0" for those who responded "heterosexual." We note that we conceptualize LGB status as a proxy for exposure to minority stress related to sexual minority identity.

<u>Moderator, gender</u>: Gender is assessed on the NSDUH as male or female[1].

<u>Outcome, past-month smoking status</u>: Binary indicator for smoking at least one cigarette in the past 30 days.

---

[1] Transgender and non-binary gender options are not provided in the NSDUH.

Demographic covariates included: age (categorized as: 18-25, 26-34, 35-49, 50+), race/ethnicity (non-Hispanic White; non-Hispanic Black; Hispanic; Asian; Native American/Alaskan Native; Native Hawaiian/Other Pacific Islander; non-Hispanic other race/multiracial), education level (less than high school; high school; some college/2-year college degree; 4-year college degree), employment (full-time; part-time; student; unemployed; other), and household income (less than $20,000; $20,000-$49,999; $50,000-$74,999; $75,000+).

**Potential outcomes, propensity scores, and moderated treatment effects**

We first introduce standard potential outcomes notation to define key estimands of interest **for a binary treatment or exposure**. Each individual has two potential outcomes: $Y_1$, their outcome if they experienced the exposure condition (e.g., LGB identity and the associated social stressors **for our case study**), and $Y_0$, their outcome if they experienced the comparison condition (e.g., heterosexual identity). $Y_1$ and $Y_0$ exist for all individuals in the population; however, we only observe one for each participant, corresponding to their reported **exposure status**. Let $T$ denote our binary **exposure of interest**. Then, $Y_i^{obs} = Y_{1i} \times T_i + Y_{0i} \times (1 - T_i)$ denotes the i[th] individual's observed outcome.

The treatment effect[2] for an individual is defined as $Y_{1i} - Y_{0i}$, representing the difference in their potential outcomes; the average treatment effect across the population (ATE) is defined as $E(Y_1 - Y_0)$. In our case study, the ATE represents the average effect of minority stress due to LGB identity on current smoking rates across the entire population. Let $Z$ represent a binary moderator (e.g., gender **in our case study**). Often, it is the case that $E(Y_1 - Y_0|Z = 1) \neq E(Y_1 - Y_0|Z = 0)$ – e.g., the ATE for minority stress due to LGB status on smoking may differ for men and women. When moderation is present, it is of interest to estimate the ATE for each level of the moderator. The moderated ATE (M-ATE) is defined as $E(Y_1 - Y_0 | Z = z)$.

---

[2] We note that some in the causal inference field have argued that treatments or exposures must be "manipulable," namely capable of being changed or randomized to individuals[46,47] in order to obtain causal effects. This viewpoint excludes investigations of the causal effects of health disparities due to race, gender, or sexual identity (i.e., generally considered non-manipulable attributes). However, our case study adopts the framework articulated by Krieger, VanderWeele, and others that causal effects can be well-defined with regard to these exposures, given the social conditions/stressors associated with being in a socially marginalized group are indeed manipulable (and have changed across time) and are a primary contributor to health outcomes.[13,48]

Our proposed method uses PS weighting to create comparable groups with respect to baseline potential confounders when estimating the M-ATE. In general, the PS is defined as an individual's probability of treatment, given their pretreatment characteristics denoted by the vector $\boldsymbol{X}$: $p(\boldsymbol{X}) = Pr(T = 1|\boldsymbol{X})$. When interest lies in estimating the M-ATE, PS estimation will additionally be conditional on the moderator,[14] as follows:

$$p(\boldsymbol{X}, Z) = Pr(T = 1|\boldsymbol{X}, Z = z)$$

PS estimates, $\hat{p}(\boldsymbol{X}, Z)$, are then used to define PS weights as follows for a binary moderator $Z$: we weight the treatment group within each level of $Z$ by $1/\hat{p}(\boldsymbol{X}_i, Z_i)$ and weight the comparison group within each level of $Z$ by $1/(1 - \hat{p}(\boldsymbol{X}_j, Z_i))$. In order to identity the ATE, and likewise the M-ATE, we must assume strong ignorability, defined as follows:

i. $(Y_1, Y_0) \perp T|X, Z$ – e.g., no unobserved confounders
ii. $0 < p(\boldsymbol{X}, Z) < 1$ – e.g., **positivity or** overlap between exposure groups within levels of the moderator

These two key assumptions (overlap and no unobserved confounders) are impossible to test in practice; however, their plausibility should be explored in any study using PS weights. This tutorial details how to assess plausibility of both the overlap assumption (in Step 1) and the no unobserved confounding assumption (in Step 5), highlighting an important and often overlooked part of many applied studies. **In the writing that follows, checking for overlap implies checking for the positivity assumption.**

We highlight that this approach can be implemented in the context of weighted data, such as data with survey sampling or attrition weights. Prior work has demonstrated that, in the context of traditional PS weighting (no moderation), it is important that the survey or attrition weights be used as survey weights in the estimation of the PS weight.[49] Then, the PS weights can be multiplied by survey or attrition weights to obtain the needed composite weights for use in the final outcome models.

**Key steps for estimating moderator treatment effects**

*Step 1: Check for covariate overlap across treatment groups for each level of the moderator*

While there is no formal way to test for overlap across treatment/exposure groups on the multivariate distribution of the control covariates, one can assess overlap in the univariate distributions of covariates. In the traditional PS context (no moderation), one should explore baseline descriptive statistics of all covariates included in the PS model, stratified by treatment group. If variables exhibit differential range across treatment groups, this suggests violation of the overlap assumption. Extending to the moderation context, covariate overlap should similarly be assessed, stratified by both the treatment and moderator. If there are obvious areas where there is a lack of overlap, it is possible to still estimate moderated effects so long as the groups for which there is a lack of overlap are removed completely from the evaluation. Removing a subsample from the data set in order to achieve overlap between the treatment groups will diminish the generalizability of study findings but is often the only option when there are obvious areas without overlap.[50]

*Step 2: Estimate the PS weights*

In the context of the M-ATE, estimation of the PS model separately within each level of the moderator generally results in better covariate balance within moderator strata which, in turn, can minimize bias.[14] However, in the case of small sample sizes, this stratified estimation approach can result in unstable weights and overly noisy estimates of the M-ATE; in these cases, alternative estimation strategies may be preferable. We address these tradeoffs more in our discussion. Since our case study data affords adequate sample size, we estimate the PS model separately for males and females. **Estimation of the PS model including selection of the most relevant covariates requires significant care; model misspecification can lead to biased results. Resultantly, machine learning methods offer a key** **advantage over parametric models – like logistic regression –in this setting.**[51] We implement PS estimation with the TWANG package[52] which uses a nonparametric machine learner, generalized boosted modeling.[53] **However, we note that alternative estimation methods could be used, as there are numerous approaches to estimating high quality PS weights, including more recent causal machine learning methods.[21,22,53-57] Additionally, variable selection requires careful consideration when estimating PS model. Selection of the covariates for use in the PS model should include variables that are associated with the outcome – often regardless of their association with the treatment indicator. However, variables that are solely predictive of treatment with no association with the outcome should be avoided.[51,58]** As noted above, we properly include the NSDUH survey weight in the estimation of the PS within each level of the moderator to ensure we obtain balance among the survey weighted version of the data.

*Step 3: Evaluate whether PS weights improved covariate balance within each level of the moderator*

In the absence of moderation, an essential diagnostic step is to evaluate the extent to which PS weighting improved covariate balance across treatment groups; **there has been rich methodological guidance on how best to do so for estimation of main effects.[51,59-61]** Extending to the moderation context, after PS weighting, covariate balance should be evaluated across treatment groups within each level of the moderator. Often, balance is assessed by strictly comparing PS weighted mean differences for each covariate across treatment groups. However, others have argued that ensuring covariate distributions are similar across treatment groups is essential for high quality PS weights.[62,63] We recommend using both standardized mean differences (SMD) and Kolmogorov-Smirnoff (KS) statistics to assess balance. The SMD is a measure of standardized mean differences between the groups, while KS statistics are measures of distributional differences between groups. More specifically, the KS statistic is the maximum absolute difference between the two empirical cumulative distribution functions of the treated and control groups for a given covariate. For both SMD and KS, smaller is better, with values of 0 indicating no differences between treatment groups. A recommended threshold is that both SMD and KS values should not exceed 0.10 after PS weighting; absolute standardized differences of 0.10 are considered to be small effect size differences[54,62,64] and recent work has shown this same cut-off point is reasonable for KS statistics.[62]

In the context of survey or attrition weighted data, comparisons prior to PS weighting should be weighted using the survey or attrition weight. Comparisons after PS weighting should use a composite weight (i.e., PS weight multiplied by the survey or attrition weight). When using TWANG, all of this is directly handled by the package.

*Step 4: Estimate moderated treatment effect*

If PS weights have successfully created comparable groups in Step 3, then moderated effects can be estimated. Specifically, M-ATE estimates will be calculated as the difference in PS weighted means between the two exposure groups (e.g., LGB and heterosexual) within each level of the moderator (e.g., females and males). This can be done by fitting a weighted regression model that includes indicators for both the treatment/exposure and the moderator, their interaction, **as well as the same set of covariates used in the PS model. This model can then be used to recover the estimated treatment effects within levels of the moderators, as described below. We suggest including all covariates used in the PS model in the final outcome model since** this "doubly robust" estimation procedure is the preferred approach for estimating treatment/exposure effects with PS weights.[16,65,66] As long as one model (either the covariate-adjusted outcome model or the PS model) is specified correctly, one will obtain consistent estimates for the ATE, and by extension, the M-ATE. Including the confounding covariates in the outcome model also reduces any residual covariate imbalance remaining after weighting and can improve precision of effect estimates. Since we are using PS weights **in our final model, we use standard survey package commands to obtain standard errors that account for the weighting**.

Recommended specification of outcome model for estimating M-ATEs:

$$g(E[Y_i^{obs}]) = \alpha_0 + \alpha_1 \times T_i + \alpha_2 \times Z_i + \alpha_3 \times T_i * Z_i + \boldsymbol{\alpha_4} \times \boldsymbol{X_i}$$

where $g(.)$ is the appropriate link function. In this model specification, we assume the absence of interactions between covariates *X* and *Z* **for illustrative purposes. If such interactions exist, it is best to control for these additional interaction terms in the final outcome model.[67] We also assume only one moderator of interest; if interest lies in estimation of multiple M-ATEs, it is possible to repeat this procedure independently for each candidate moderator.**

After estimating the outcome model, we can use post hoc estimation procedures to calculate the needed M-ATE for each level of the moderator. **For example, in our case study, smoking status is binary so $g(.)$ will be the logit link for a logistic outcome model. To obtain estimates of the M-ATE, we must use the fitted logistic regression model to estimate the marginal probabilities of the outcome for each subgroup under both exposure conditions, which can then be used to compute the needed differences in the means of the potential outcomes and risk differences for males and females.[16] That is, for every person in the sample, we estimate the probability of smoking setting sexual identity to LGB and then repeat this estimation setting sexual identity to heterosexual. For both females and males, we separately average the estimated probabilities for both LGB and heterosexual sexual identity, the difference of which (within gender groups) provides our treatment effect estimate. Averages with groups should be weighted by sampling, nonresponse or attrition weights. We report the estimates and their corresponding 95% confidence intervals.** In many studies, primary interest for the moderation analyses might focus on whether there is any evidence of moderation (rather than reporting of subgroup effects). This can be done using the outcome model above by testing the null hypothesis that $\alpha_3=0$ and can be examined directly from the regression output by referring to the 95% confidence interval for $\hat{a}_3$. When the moderator has multiple levels, then a joint test for all the interactions terms between treatment/exposure and the interaction can be used.

It would also be reasonable to estimate stratified outcome models using the PS weights from each level of the moderator and to report the estimated effects from those models. Here, we focus on estimation of the joint outcome model given it allows for researchers to directly test for moderation more formally than stratified models.

*Step 5: Check for sensitivity to unobserved confounding*

Given the strong ignorability assumption, it is critical to assess the potential impact that unobserved confounders might have on the study's key findings in any analysis using PS weights. There are many tools available to examine the potential impact of unobserved confounding for main effects of treatment.[68-72] Conveniently, most of these tools can be applied to examine sensitivity in moderation analyses as well, though with some adaptation. Here, we use a graphical tool (the Ovtool in R) that relies on using simulations to assess how sensitive our findings may be to potential unobserved **confounding by a variable that is independent of the observed confounders conditional on the treatment indicator. This graphical approach tests sensitivity** both in terms of how the estimated treatment effect and statistical significance (i.e., p-value) might change as a function of the strength of the relationships between the unobserved confounders and the treatment/exposure and outcome within each level of the moderator.[68,73]

## Illustrative Case Study Application

Table 1 shows the comparison of LGB and heterosexual adults, stratified by gender, prior to PS weighting. SMDs and KS statistics indicate clear differences between heterosexual and LGB individuals for both males and females. In general, LGB adults are more likely to be unemployed, are younger, and have lower household income than heterosexual adults. The magnitude of the difference between LGB and heterosexual adults varies by gender – e.g., age differences are more disparate for females than males. While there are no notable differences between female LGB and heterosexual females on race/ethnicity, heterosexual males are less likely to be Hispanic than LGB males. LGB females are less likely than heterosexual females to be college graduates, while there are no such noted differences for males.

*Step.1 Check for covariate overlap across treatment groups for each level of the moderator*

**Given the categorical nature of our covariates, overlap concerns can be assessed by checking whether there are subgroups within each covariate that have no individuals after stratifying by both LGB identity and gender (see Table 1). Having no individuals in one specific subgroup (e.g., less than high school) would imply that the analysis cannot estimate the causal effect of the treatment for that specific subgroup. Thus, systematically checking the covariate distribution via frequency tables can be a pragmatic way to identify lack of overlap (when present) for categorical variables. As no cells with 0% are noted, there does not appear to be any concerns regarding covariate overlap for this analysis. For continuous covariates, we recommend comparing the minimum and maximum for each treatment group within each level of the moderator; histograms stratified with respect to the moderator can also be examined.**

*Step.2 Estimate the PS weights*

We estimate the PS model separately for males and females using the TWANG package. Code for STATA and R are provided in the Appendix. The NSDUH survey weight is included via the *sampw* option in TWANG **which uses the weights in estimating the propensity score models and the post hoc checks for balance in the weighted covariate distributions.**

*Step.3 Evaluate whether PS weights improved covariate balance within each level of the moderator*

After PS weighting, we evaluate covariate balance with respect to both SMDs and KS statistics separately for both females and males (see Figure 1, Table 2). We note that balance is assessed here using the composite weight (i.e., PS weight multiplied by the survey weight), calculated by TWANG. All absolute SMDs and KS statistics fall well below the 0.10 threshold, suggesting successful balance has been achieved within each level of the moderator.

*Step.4 Estimating moderated treatment effects*

Next, we obtain M-ATE estimates using covariate-adjusted weighted logistic regression. As in Step 3, we use the composite weights in the outcome model. We observe significant evidence that LGB status is associated with higher likelihood of smoking for both males and females (Table 3). Using the marginal probabilities from the regression results, the estimated M-ATE risk difference of current smoking for LGB vs heterosexual males is 0.05 (95% CI = 0.00, 0.10), suggesting LGB males are moderately more likely to smoke than their heterosexual counterparts. In contrast, LGB females have substantially higher risk of smoking than heterosexual females, with an estimated M-ATE risk difference of 0.15 (95% CI = 0.08, 0.18). There is significant evidence of moderation, as determined by the significant coefficient for the interaction term **in the logistic regression model** – namely 0.65 (95% CI = 0.23, 1.01).

*Step 5. Assess sensitivity to unobserved confounding*

Finally, we conduct sensitivity analyses for unobserved confounding for the estimated M-ATE effects separately for females and males. Figure 2 shows how both the estimated effect (solid contours) and p-value (dashed contours) would change as a function of an unobserved confounder whose association **with our key exposure (LGB status)** is expressed through an effect size or SMD (x-axis), and whose relationship with the outcome is expressed as a correlation (y-axis). The solid contours report the adjusted M-ATE estimates that would result as the relationship between the omitted variable and the outcome and exposure group indicator increases **(LGB vs not).**

**As shown for females, the estimated treatment effect gets larger as we move to the left from zero along the x-axis – going from 0.2 to 0.4 as we move to the upper left hand corner - and decreases as we move right from zero – crossing over a null effect of 0 as we move to the upper right hand corner. The dashed contours show how statistical significance of our result is impacted and parallel what we see for the estimated treatment effect contours. Namely, there is an area of the plot around the 0 treatment effect contour that indicates that statistically significance disappears with certain magnitudes of an unobserved confounder. For females, an unobserved confounder would need a correlation with the outcome greater than 0.10 and would need to differ between LGB and heterosexual groups with an SMD of greater than 0.42 to change our finding such that it is no**

**longer statistically significant (at 0.05 level). Additionally, the plot shows (via the black dots) the observed correlations and effect sizes observed in the data for the different categories of the five observed covariates used in the PS weights. As shown, the observed covariates are far from the region of the plot where the unobserved covariates would need to be to impact our M-ATE estimate for females. The vast majority of the observed covariates are not correlated with the outcome at greater than 0.10 and have unweighted SMDs between exposure groups of less than 0.42.**

**Thus, relative to the observed covariates, the omitted variable would need to have a very strong relationship with both the outcome and the treatment group in order to change our findings regarding the observed association between LGB status and a higher likelihood of smoking for females. This sensitivity analysis suggests that if omitted variables are present, our findings for females would be unlikely to substantially change if these variables were controlled for, indicating that our results are robust to unobserved confounding.**

**In contrast, the results are notably different for males. We see evidence of a highly sensitive finding that could easily move towards the null with the addition of even a weakly correlated omitted variable, suggesting the finding for males is not as sufficiently trustworthy and should be interpreted cautiously as it could be the result of an omitted variable, unlike the more robust findings for females.**

## Discussion

This tutorial provides health researchers with guidance on how to use PS weighting for estimation of moderated effects. It illustrates the key steps using an application examining the potential association between minority stress due to LGB status and smoking among adults. Critically, we detail how to assess key assumptions required for robust estimation – including attention to overlap and unobserved confounding sensitivity analyses. Similar to prior work on PS matching, we recommend that estimation of the PS weights should occur within each level of the moderator when feasible.[14] This tutorial provides guidance for performing moderation using PS weighting and ensures careful consideration of key assumptions (namely overlap and unobserved confounding) that is often overlooked in practice.

We highlight several additional methodological considerations in the context of moderation analysis. First, there are a wealth of methods available for healthcare researchers to examine treatment effect heterogeneity and the choice of the best method for a given case study should be driven by the scientific goals of the research. **While we utilize the TWANG package to estimate our PS weights, our steps generalize beyond the use of this method for estimation of the PS weights. Researchers can find an overview to the numerous causal machine learner methods in this space as well as python code[74] using the GitHub page from Microsoft Research's Automated Learning and Intelligence for Causation and Economics (ALICE) project.** Additionally, we examine a binary moderator; alternative methods are needed to address continuous moderator variables. Finally, caution is necessary when using PS weighting for moderation analyses with smaller sample sizes. While PS weighting within each level of the moderator can minimize M-ATE bias arising from imbalance within moderator subgroups, it comes at the expense of more variable weights, leading to larger standard errors for M-ATE estimates. In some studies, this loss in precision may come at too great a cost and fitting a single PS model (pooling across moderation levels) may be preferrable. There is no simple rule of thumb for selecting among these options but comparing the balance achieved and associated loss in precision due to each option can help analysts to choose the optimal approach for a given study.

In addition, we highlight that there is a growing literature on weighting methods that deal with the lack of overlap issue, including the use of overlap weighting.[75,76] These methods target the subset of the population for which data supports overlap in baseline characteristics. That is, they estimate an average treatment effect on the overlap sample, denoted ATO. Overlap weighting methods offer an ability to obtain good balance and optimal variance on the sample where overlap occurs. However, it is not always readily apparent if the ATO is of substantive interest, though many argue this is the subset where researchers have clinical equipoise. Notably, in the context of moderation analyses, we highlight that the population subset with appropriate overlap could potentially differ across moderator levels. Future work should seek to generalize overlap weighting techniques for moderation analysis; for example, identifying the population subset which has adequate overlap for all moderator levels may be of interest such that M-ATEs are directly comparable across moderator groups. **We note that one of the goals of this tutorial is to provide researchers with a way to check for violations of overlap prior to estimation of the PS weights. More traditionally, it has been recommended that one assess reasonableness of the overlap assumption post hoc by comparing the estimated PS distributions after estimating the weights.[51,60] Unfortunately, such a post hoc comparison alone does not allow researchers to see what might be driving the regions where there is a lack of overlap.**

Overall, estimating robust moderation effects requires thoughtful consideration of covariate imbalances across both treatment and moderator levels; as detailed in this tutorial, PS weighting is one promising technique to minimize bias arising from covariate imbalances.

# References

1. Ehde DM, Arewasikporn A, Alschuler KN, et al. Moderators of treatment outcomes after telehealth self-management and education in adults with multiple sclerosis: A secondary analysis of a randomized controlled trial. Arch Phys Med Rehabil. 2018;99(7):1265-1272.

2. Falk DE, Castle IJP, Ryan M, et al. Moderators of varenicline treatment effects in a double-blind, placebo-controlled trial for alcohol dependence: An exploratory analysis. J Addict Med. 2015;9(4):296-303.

3. Kraemer HC, Wilson GT, Fairburn CG, et al. Mediators and moderators of treatment effects in randomized clinical trials. Arch Gen Psychiatry. 2002;59(10):877-883.

4. Fernandez y Garcia E, Nguyen H, Duan N, et al. Assessing heterogeneity of treatment effects: Are authors misinterpreting their results? Health Serv Res. 2010;45(1):283-301.

5. Lazar AA, Bonetti M, Cole BF, et al. Identifying treatment effect heterogeneity in clinical trials using subpopulations of events: STEPP. Clin Trials. 2016;13(2):169-179.

6. Varadhan R, Seeger JD. Chapter 3: Estimation and Reporting of Heterogeneity of Treatment Effects. In: Velentgas P, Dreyer N, Nourjah P, et al., eds. Developing a Protocol for Observational Comparative Effectiveness Research: A User's Guide. Rockville, MD: Agency for Healthcare Research and Quality; 2013.

7. Rosenkranz G. Chapter 2: Objectives and Current Practice of Subgroup Analyses. In: Rosenkranz G, ed. Exploratory Subgroup Analyses in Clinical Research.2019:31-45.

8. Paget M-A, Chuang-Stein C, Fletcher C, et al. Subgroup analyses of clinical effectiveness to support health technology assessments. Pharmaceutical Statistics. 2011;10(6):532-538.

9. Sturmer T, Rothman KJ, Glynn RJ. Insights into different results from different causal contrasts in the presence of effect-measure modification. Pharmacoepidemiol Drug Saf. 2006;15(10):698-709.

10. Kurth T, Walker AM, Glynn RJ, et al. Results of multivariable logistic regression, propensity matching, propensity adjustment, and propensity-based weighting under conditions of nonuniform effect. Am J Epidemiol. 2006;163(3):262-270.

11. VanderWeele TJ. On the distinction between interaction and effect modification. Epidemiology. 2009;20(6):863-871.

12. Lunt M, Solomon D, Rothman K, et al. Different methods of balancing covariates leading to different effect estimates in the presence of effect modification. Am J Epidemiol. 2009;169(7):909-917.

13. VanderWeele TJ, Hernan MA. Causal effects and natural laws: Towards a conceptualization of causal counterfactuals for nonmanipulable exposures, with application to the effects of race and sex. In: Berzuini C, Dawid P, L B, eds. Causality: Statistical Perspectives and Applications. New York, NY: Wiley; 2012:101–113.

14. Green KM, Stuart EA. Examining moderation analyses in propensity score methods: Application to depression and substance use. J Consult Clin Psychol. 2014;82(5):773-783.

15. Chang T-H, Stuart EA. Overview of methods for adjustment and applications in the social and behavioral sciences: The role of study design. In: Handbook of Matching and Weighting Adjustments for Causal Inference. Chapman and Hall/CRC; 2023:3-20.

16. Bang H, Robins JM. Doubly robust estimation in missing data and causal inference models. Biometrics. 2005;61(4):962-972.

17. Stuart EA. Matching methods for causal inference: A review and a look forward. Stat Sci. 2010;25(1):1-21.

18. Kong Y, Zhou J, Zheng Z, et al. Using machine learning to advance disparities research: Subgroup analyses of access to opioid treatment. Health Serv Res. 2021.

19. Xu Y, Yu M, Zhao Y-Q, et al. Regularized outcome weighted subgroup identification for differential treatment effects. Biometrics. 2015;71(3):645-653.

20. Yuan A, Chen X, Zhou Y, et al. Subgroup analysis with semiparametric models toward precision medicine. Stat Med. 2018;37(11):1830-1845.

21. Athey S, Imbens G. Recursive partitioning for heterogeneous causal effects. Proc Natl Acad Sci U S A. 2016;113(27):7353-7360.

22. Wager S, Athey S. Estimation and inference of heterogeneous treatment effects using random forests. J Am Stat Assoc. 2018;113(523):1228-1242.

23. Athey S, Wager S. Policy learning with observational data. Econometrica. 2021;89(1):133-161.

24. Nie X, Wager S. Quasi-oracle estimation of heterogeneous treatment effects. Biometrika. 2021;108(2):229-319.

25. Syrgkanis V, Lei V, Oprescu M, et al. Machine learning estimation of heterogeneous treatment effects with instruments. Proceedings of the 33rd Conference on Neural Information Processing Systems (NeurIPS). 2019.

26. Foster D, Syrgkanis V. Orthogonal statistical learning. Proceedings of the 32nd Annual Conference on Learning Theory (COLT). 2019.

27. Oprescu M, Syrgkanis V, Wu Z. Orthogonal random forest for causal inference. Proceedings of the 36th International Conference on Machine Learning (ICML). 2019.

28. Kunzel S, Sekhon J, Bickel J, et al. Metalearners for estimating heterogeneous treatment effects using machine learning. Proceedings of the National Academy of Sciences. 2019;116(10):4156-4165.

29. Athey S, Tibshirani J, Wager S. Generalized random forests. Ann Stat. 2019;47(2):1148-1178.

30. Chernozhukov V, Nekipelov D, Semenova V, et al. Plug-in regularized estimation of high-dimensional parameters in nonlinear semiparametric models. Arxiv preprint arxiv:1806.04823. 2018.

31. Chernozhukov V, Chetverikov D, Demirer M, et al. Double machine learning for treatment and causal parameters. ArXiv preprint arXiv:1608.00060. 2016.

32. Dudik M, Hehan D, Langford J, et al. Doubly robust policy evaluation and optimization. Stat Sci. 2014;29(4):485-511.

33. Chernozhukov V, Chetverikov D, Demirer M, et al. Double/debiased machine learning for treatment and structural parameters. Econometrics Journal. 2018;21(1):C1-C68.

34. Schuler MS, Rice CE, Evans-Polce RJ, et al. Disparities in substance use behaviors and disorders among adult sexual minorities by age, gender, and sexual identity. Drug Alcohol Depend. 2018;189:139-146.

35. Gonzales G, Przedworski J, Henning-Smith C. Comparison of health and health risk factors between lesbian, gay, and bisexual adults and heterosexual adults in the United States: Results from the National Health Interview Survey. JAMA Intern Med. 2016;176(9):1344-1351.

36. Gordon AR, Fish JN, Kiekens WJ, et al. Cigarette smoking and minority stress across age cohorts in a national sample of sexual minorities: Results from the Generations Study. Ann Behav Med. 2021;55(6):530-542.

37. Mereish EH, Goldbach JT, Burgess C, et al. Sexual orientation, minority stress, social norms, and substance use among racially diverse adolescents. Drug Alcohol Depend. 2017;178:49-56.

38. Lee JH, Gamarel KE, Bryant KJ, et al. Discrimination, mental health, and substance use disorders among sexual minority populations. LGBT Health. 2016;3(4):258-265.

39. Goldbach JT, Tanner-Smith EE, Bagwell M, et al. Minority stress and substance use in sexual minority adolescents: A meta-analysis. Prev Sci. 2014;15(3):350-363.

40. Meyer IH. Prejudice, social stress, and mental health in lesbian, gay, and bisexual populations: Conceptual issues and research evidence. Psychol Bull. 2003;129(5):674-697.

41. Meyer JP, Isaacs K, El-Shahawy O, et al. Research on women with substance use disorders: Reviewing progress and developing a research and implementation roadmap. Drug Alcohol Depend. 2019;197:158-163.

42. McHugh RK, Votaw VR, Sugarman DE, et al. Sex and gender differences in substance use disorders. Clinical Psychology Review. 2018;66:12-23.

43. Schuler MS, Stein BD, Collins RL. Differences in substance use disparities across age groups in a national cross-sectional survey of lesbian, gay, and bisexual adults. LGBT Health. 2019;6(2):68-76.

44. Hoffman L, Delahanty J, Johnson SE, et al. Sexual and gender minority cigarette smoking disparities: An analysis of 2016 Behavioral Risk Factor Surveillance System data. Prev Med. 2018;113:109-115.

45. SAMHSA. *2018 National Survey on Drug Use and Health: Methodological summary and definitions.* Rockville, MD: Center for Behavioral Health Statistics and Quality, Substance Abuse and Mental Health Services Administration;2019.

46. Imbens G, Rubin D. Causal Inference for Statistics, Social and Biomedical Sciences: An Introduction. Cambridge, UK: Cambridge University Press; 2015.

47. Holland P. Statistics and causal inference. J Am Stat Assoc. 1986;81:945–960.

48. Krieger N, Davey Smith G. The tale wagged by the DAG: Broadening the scope of causal inference and explanation for epidemiology. Int J Epidemiol. 2016;45(6):1787-1808.

49. Ridgeway G, Kovalchik SA, Griffin BA, et al. Propensity score analysis with survey weighted data. J Causal Inference. 2015;3(2):237-249.

50. Crump RK, Hotz VJ, Imbens GW, et al. Dealing with limited overlap in estimation of average treatment effects. Biometrika. 2009;96(1):187-199.

51. Desai RJ, Franklin JM. Alternative approaches for confounding adjustment in observational studies using weighting based on the propensity score: a primer for practitioners. BMJ. 2019;367:l5657.

52. twang: Toolkit for Weighting and Analysis of Nonequivalent Groups. https://cran.r-project.org/web/packages/twang/index.html. Accessed March 15, 2022.

53. McCaffrey DF, Ridgeway G, Morral AR. Propensity score estimation with boosted regression for evaluating causal effects in observational studies. Psychol Methods. 2004;9(4):403-425.

54. Austin PC, Stuart EA. Moving towards best practice when using inverse probability of treatment weighting (IPTW) using the propensity score to estimate causal treatment effects in observational studies. Stat Med. 2015;34(28):3661-3679.

55. Imai K, Ratkovic M. Covariate balancing propensity score. Journal of the Royal Statistical Society: Series B (Statistical Methodology). 2014;76(1):243-263.

56. Hainmueller J. Entropy balancing for causal effects: A multivariate reweighting method to produce balanced samples in observational studies. Political Analysis. 2012;20(1):25-46.

57. Zubizarreta JR. Stable weights that balance covariates for estimation with incomplete outcome data. J Am Stat Assoc. 2015;110(511):910-922.

58. Myers JA, Rassen JA, Gagne JJ, et al. Effects of adjusting for instrumental variables on bias and precision of effect estimates. Am J Epidemiol. 2011;174(11):1213-1222.

59. Williamson E, Morley R, Lucas A, et al. Propensity scores: from naive enthusiasm to intuitive understanding. Stat Methods Med Res. 2012;21(3):273-293.

60. Webster-Clark M, Sturmer T, Wang T, et al. Using propensity scores to estimate effects of treatment initiation decisions: State of the science. Stat Med. 2021;40(7):1718-1735.

61. D'Agostino RB, Jr. Propensity score methods for bias reduction in the comparison of a treatment to a non-randomized control group. Stat Med. 1998;17(19):2265-2281.

62. Markoulidakis A, Holmans P, Pallmann P, et al. How balance and sample size impact bias in the estimation of causal treatment effects: A simulation study. arXiv:2107.09009. arXiv. 2021.

63. Zhou T, Tong G, Li F, et al. PSweight: An R package for propensity score weighting analysis. R Journal. 2022;14:282-300.

64. Austin PC. Balance diagnostics for comparing the distribution of baseline covariates between treatment groups in propensity-score matched samples. Stat Med. 2009;28(25):3083-3107.

65. Chattopadhyay A, Hase CH, Zubizarreta JR. Balancing vs modeling approaches to weighting in practice. Stat Med. 2020;39(24):3227-3254.

66. Kang JDY, Schafer JL. Demystifying double robustness: A comparison of alternative strategies for estimating a population mean from incomplete data. Stat Sci. 2007;22(4):523-539.

67. Marsden AM, Dixon WG, Dunn G, et al. The impact of moderator by confounder interactions in the assessment of treatment effect modification: a simulation study. BMC Med Res Methodol. 2022;22(1):88.

68. Burgette L, Griffin BA, Pane J, et al. Got a feeling something is missing? Assessing Sensitivity to Omitted Variables. To be submitted.

69. Arah OA, Chiba Y, Greenland S. Bias formulas for external adjustment and sensitivity analysis of unmeasured confounders. Ann Epidemiol. 2008;18(8):637-646.

70. McCandless LC, Gustafson P, Levy A. Bayesian sensitivity analysis for unmeasured confounding in observational studies. Stat Med. 2007;26(11):2331-2347.

71. Stürmer T, Schneeweiss S, Avorn J, et al. Adjusting effect estimates for unmeasured confounding with validation data using propensity score calibration. Am J Epidemiol. 2005;162(3):279-289.

72. VanderWeele TJ, Ding P. Sensitivity analysis in observational research: Introducing the E-Value. Ann Intern Med. 2017;167(4):268-274.

73. OVtool: Omitted Variable Tool. https://cran.r-project.org/web/packages/OVtool/index.html. Accessed March 15, 2022.

74. PyWhy. https://github.com/py-why/EconML. Published 2022. Updated March 1, 2023. Accessed May 3, 2023.

75. Li F, Thomas LE, Li F. Addressing extreme propensity scores via the overlap weights. Am J Epidemiol. 2019;188(1):250-257.

76. Li F, Lock Morgan K, Zaslavsky A. Balancing covariates via propensity score weighting. J Am Stat Assoc. 2018;113(521):390-400.

**Figure Legends**

**Figure 1. SMDs before and after weighting for Females and Males**

**Figure 2. Results from omitted variable sensitivity analyses**

# Figures

**Figure 1. SMDs before and after weighting for Females and Males.**

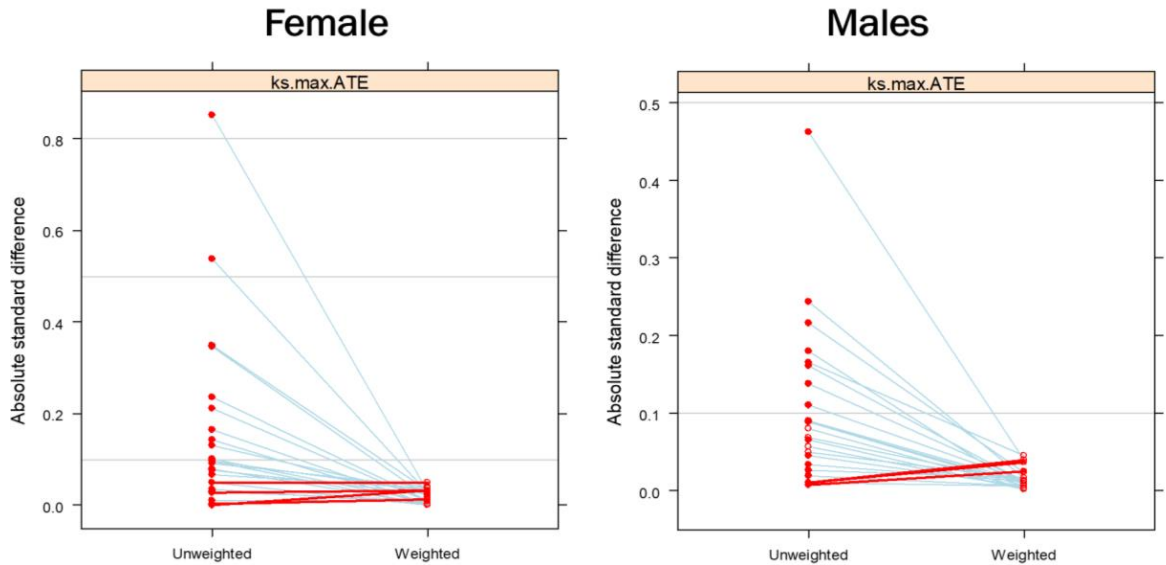

**\* Note: "ks.max.ATE" in the figure refers to the stopping rule used in TWANG and not KS statistics directly. In TWANG, a user can select from 4 stopping rules when estimating the propensity score weights: namely rules that aim to optimize (here minimize) the mean or max SMD or the mean and max KS statistics.**

**Figure 2. Results from omitted variable sensitivity analyses**

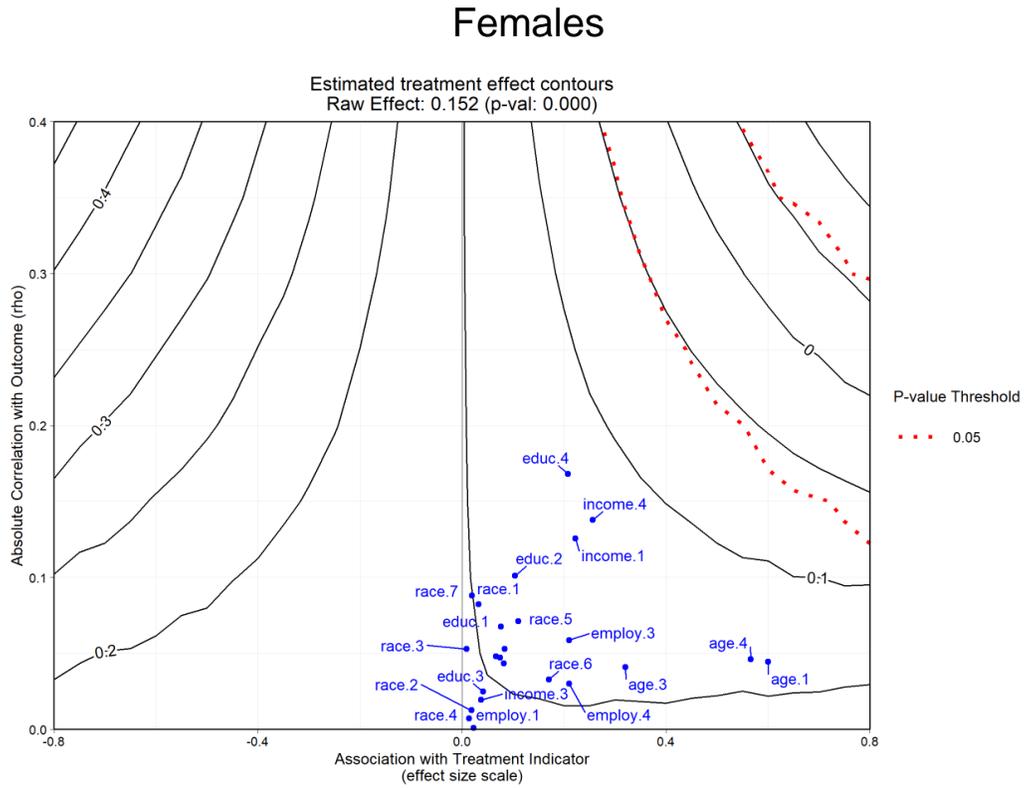

Females

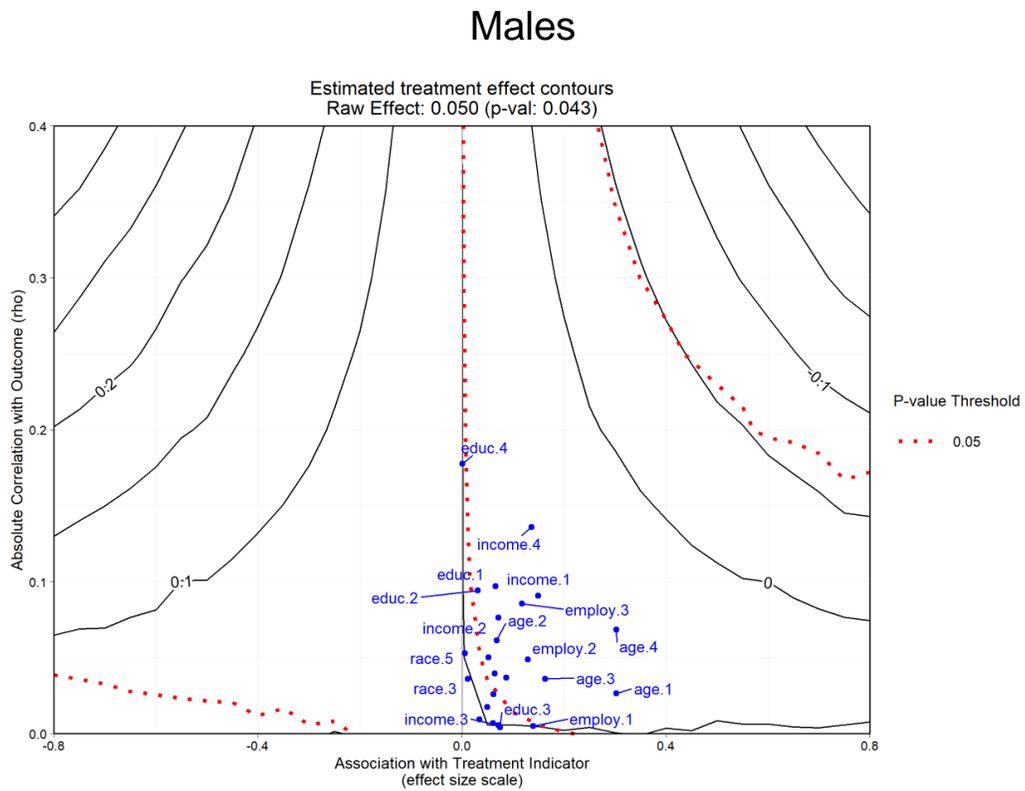

Males

Table 1Table 1

# Tables

**Table 1. Baseline characteristics of LGB and heterosexual adults prior to PS weighting, by gender**

|  | Females | | Males | |
|---|---|---|---|---|
|  | LGB | Heterosexual | LGB | Heterosexual |
| **Age** | | | | |
| 18-25 Years Old | 36.3% | 11.2%[a,b] | 23.9% | 13.5%[a] |
| 26-34 Years Old | 28.5% | 14.5%[a,b] | 25.8% | 16.1%[a] |
| 35-49 Years Old | 19.9% | 24.2%[a] | 23.3% | 24.8% |
| 50 or Older | 15.3% | 50%[a,b] | 27.1% | 45.6%[a,b] |
| **Race/Ethnicity** | | | | |
| NonHisp White | 59.8% | 63.6% | 59.8% | 64.2% |
| NonHisp Black/Afr Am | 14.2% | 12.5% | 9.7% | 11.2% |
| NonHisp Nat.Am/AK.Nat | 0.6% | 0.6% | 0.4% | 0.5% |
| NonHisp Nat.HI/Other PIs | 0.3% | 0.3% | 0.6% | 0.4% |
| NonHisp Asian | 3.8% | 5.7% | 4.8% | 5.4% |
| NonHisp multiracial/other race | 4.2% | 1.6%[a] | 1.9% | 1.7% |
| Hispanic | 17.0% | 15.7% | 22.7% | 16.5%[a] |
| **Education** | | | | |
| Less high school | 12.5% | 10.9% | 9.7% | 12.4% |
| High school grad | 23.6% | 22.4% | 23.7% | 26.3% |
| Some college/Assoc Dg | 36.1% | 32.8% | 31.0% | 28.7% |
| College graduate | 27.9% | 33.8%[a] | 35.6% | 32.6% |
| **Income** | | | | |
| Less than $20,000 | 24.2% | 15.8%[a] | 17.4% | 12.4%[a] |
| $20,000 - $49,999 | 30.5% | 30.0% | 30.4% | 26.3% |
| $50,000 - $74,999 | 14.8% | 15.9% | 15.8% | 16.2% |
| $75,000 or More | 30.5% | 38.3%[a] | 36.4% | 45.2%[a] |
| **Employment** | | | | |
| Employed full time | 46.8% | 42.0% | 57.3% | 57.9% |
| Employed part time | 19.4% | 15.4% | 13.0% | 10.1% |
| Unemployed | 8.4% | 3.2%[a] | 7.0% | 4.2%[a] |
| Student | 22.2% | 37.6%[a,b] | 19.8% | 26.1%[a] |
| Other | 3.2% | 1.7% | 2.9% | 1.7% |

Notes: [a] = Denotes absolute SMD > 0.1; [b] = Denotes KS-statistic is > 0.1 All descriptive statistics account for NSDUH survey weights

**Table 2. Baseline characteristics of LGB and heterosexual adults after PS weighting, by gender**

|  | Females | | | | Males | | | |
| --- | --- | --- | --- | --- | --- | --- | --- | --- |
|  | LGB | Hetero-sexual | SMD | KS | LGB | Hetero-sexual | SMD | KS |
| **Age** | | | | | | | | |
| 18-25 Years Old | 14.0% | 13.1% | 0.03 | 0.01 | 14.8% | 14.0% | 0.02 | 0.01 |
| 26-34 Years Old | 16.1% | 15.5% | 0.02 | 0.01 | 17.2% | 16.6% | 0.02 | 0.01 |
| 35-49 Years Old | 23.9% | 23.9% | 0.00 | 0.00 | 25.2% | 24.7% | 0.01 | 0.01 |
| 50 or Older | 46.0% | 47.5% | -0.03 | 0.02 | 42.8% | 44.8% | -0.04 | 0.02 |
| **Race/Ethnicity** | | | | | | | | |
| NonHisp White | 64.5% | 63.3% | 0.03 | 0.01 | 64.4% | 64.0% | 0.01 | 0.00 |
| NonHisp Black/Afr Am | 12.6% | 12.6% | 0.00 | 0.00 | 11.3% | 11.2% | 0.01 | 0.00 |
| NonHisp Nat.Am/AK.Nat | 0.5% | 0.6% | -0.01 | 0.00 | 0.3% | 0.5% | -0.03 | 0.00 |
| NonHisp Nat.HI/OtherPIs | 0.1% | 0.3% | -0.03 | 0.00 | 0.4% | 0.5% | -0.01 | 0.00 |
| NonHisp Asian | 4.7% | 5.6% | -0.04 | 0.01 | 5.0% | 5.4% | -0.02 | 0.00 |
| NonHisp multiracial/other race | 1.8% | 1.8% | 0.00 | 0.00 | 1.3% | 1.7% | -0.04 | 0.01 |
| Hispanic | 15.9% | 15.8% | 0.00 | 0.00 | 17.3% | 16.8% | 0.01 | 0.01 |
| **Education** | | | | | | | | |
| Less high school | 9.4% | 11.0% | -0.05 | 0.02 | 12.4% | 12.3% | 0.00 | 0.00 |
| High school grad | 21.1% | 22.5% | -0.03 | 0.01 | 26.3% | 26.2% | 0.00 | 0.00 |
| Some college/Assoc Dg | 34.9% | 33.1% | 0.04 | 0.02 | 28.0% | 28.8% | -0.02 | 0.01 |
| College graduate | 34.5% | 33.4% | 0.02 | 0.01 | 33.3% | 32.8% | 0.01 | 0.01 |
| **Income** | | | | | | | | |
| Less than $20,000 | 16.7% | 16.4% | 0.01 | 0.00 | 13.0% | 12.6% | 0.01 | 0.00 |
| $20,000 - $49,999 | 29.6% | 30.1% | -0.01 | 0.01 | 25.9% | 26.5% | -0.01 | 0.01 |
| $50,000 - $74,999 | 16.2% | 15.9% | 0.01 | 0.00 | 16.4% | 16.2% | 0.01 | 0.00 |
| $75,000 or More | 37.5% | 37.7% | 0.00 | 0.00 | 44.7% | 44.8% | 0.00 | 0.00 |
| **Employment** | | | | | | | | |
| Employed full time | 44.0% | 42.4% | 0.03 | 0.02 | 59.8% | 57.8% | 0.04 | 0.02 |
| Employed part time | 15.4% | 15.7% | -0.01 | 0.00 | 10.0% | 10.2% | -0.01 | 0.00 |
| Unemployed | 3.8% | 3.6% | 0.01 | 0.00 | 4.6% | 4.4% | 0.01 | 0.00 |
| Student | 34.9% | 36.5% | -0.03 | 0.02 | 23.9% | 25.8% | -0.05 | 0.02 |
| Other | 1.9% | 1.8% | 0.00 | 0.00 | 1.7% | 1.8% | 0.00 | 0.00 |

Notes: [a] = Denotes absolute SMD > 0.1; [b] = Denotes KS-statistic is > 0.1 All descriptive statistics account for NSDUH survey weights.

**Table 3. Logistic regression results for the PS weighted doubly robust logistic model**

|  | Regression Coefficient | Standard Error | 95% Confidence Interval |
|---|---|---|---|
| (Intercept) | -0.40 | 0.19 | **(-0.77, -0.03)** |
| LGB Status | 0.31 | 0.15 | **(0.02, 0.60)** |
| Female | -0.29 | 0.04 | **(-0.37, -0.21)** |
| LGB Status*Female | 0.65 | 0.20 | **(0.26, 1.04)** |
| Age (reference = 18 – 25) |  |  |  |
| 26-34 Years Old | 0.68 | 0.09 | **(0.50, 0.86)** |
| 35-49 Years Old | 0.71 | 0.10 | **(0.51, 0.91)** |
| 50 or Older | 0.25 | 0.15 | **(-0.04, 0.54)** |
| Race/ethnic (ref. = NonHisp White) |  |  |  |
| NonHisp Black/Afr Am | -0.31 | 0.17 | **(-0.64, 0.02)** |
| NonHisp Nat.Am/AK.Nat | 0.05 | 0.29 | **(-0.52, 0.62)** |
| NonHisp Nat.HI/OtherPIs | -0.11 | 0.52 | **(-1.13, 0.91)** |
| NonHisp Asian | -0.29 | 0.46 | **(-1.19, 0.61)** |
| NonHisp >1 race | 0.33 | 0.21 | **(-0.08, 0.74)** |
| Hispanic | -0.87 | 0.14 | **(-1.14, -0.60)** |
| Education (ref. = < high school) |  |  |  |
| High school grad | 0.01 | 0.18 | **(-0.34, 0.36)** |
| Some coll/Assoc Dg | -0.34 | 0.17 | **(-0.67, -0.01)** |
| College graduate | -1.27 | 0.20 | **(-1.66, -0.88)** |
| Income (reference = <$20,000) |  |  |  |
| $20,000 - $49,999 | -0.48 | 0.15 | **(-0.77, -0.19)** |
| $50,000 - $74,999 | -0.78 | 0.17 | **(-1.11, -0.45)** |
| $75,000 or More | -1.02 | 0.17 | **(-1.35, -0.69)** |
| Employment (reference = Full time) |  |  |  |
| Employed part time | -0.42 | 0.16 | **(-0.73, -0.11)** |
| Unemployed | 0.17 | 0.18 | **(-0.18, 0.52)** |
| Student | -0.38 | 0.14 | **(-0.65, -0.11)** |
| Other | -0.75 | 0.21 | **(-1.16, -0.34)** |

**Supplementary code for both our illustrative case study as well as a generic data set**

Appendix A: Code to run for our illustrative case study data

Please note that case study data also available as part of this paper's supplementary material

**R Code**

```
#This is the R code to run the needed steps in the tutorial
#Uses the NSDUH case study which can be downloaded from our supplementary data files

library(twang)

#Subset data down to the different levels of the moderator

data.female=data[data$female==1,]
data.male=data[data$female==0,]

##########################################
#Step 1 - check for overlap concerns
##########################################

#Since everything is categorical - we simply need to check for empty cells
table(data.female$lgb_flag,data.female$age)
table(data.female$lgb_flag,data.female$race)
table(data.female$lgb_flag,data.female$educ)
table(data.female$lgb_flag,data.female$income)
table(data.female$lgb_flag,data.female$employ)

table(data.male$lgb_flag,data.male$age)
table(data.male$lgb_flag,data.male$race)
table(data.male$lgb_flag,data.male$educ)
table(data.male$lgb_flag,data.male$income)
table(data.male$lgb_flag,data.male$employ)

#If there was a continuous covariate, check the minimum and maximum for each level of the moderator
#summary(data.female$continuous.covariate)
#summary(data.male$continuous.covariate)

##################################################################################
#Step 2 - estimate PS weights using TWANG within levels of the moderator
##################################################################################

set.seed(175659)

#Note that NSDUH sampling weights are controlled for by using the sampw option

ps1.f <- ps(lgb_flag ~ age+race+educ+income+employ, data = data.female,
       n.trees=10000,interaction.depth=2,shrinkage=0.01,perm.test.iters=0,
       stop.method=c("ks.max"),estimand = "ATE",verbose=FALSE,sampw = data.female$analwt_c)

ps1.m <- ps(lgb_flag ~ age+race+educ+income+employ, data = data.male,
       n.trees=10000,interaction.depth=2,shrinkage=0.01,perm.test.iters=0,
       stop.method=c("ks.max"),estimand = "ATE",verbose=FALSE,sampw = data.male$analwt_c)
```

```r
################################################################
#Step 3 - assess balance within levels of the moderator
################################################################

plot(ps1.f)
balance1.f <- bal.table(ps1.f)
balance1.f
summary(ps1.f)

plot(ps1.m)
balance1.m <- bal.table(ps1.m)
balance1.m
summary(ps1.m)

plot(ps1.f,plots=1)
plot(ps1.m,plots=1)

plot(ps1.f,plots=2)
plot(ps1.m,plots=2)

plot(ps1.f,plots=3)
plot(ps1.m,plots=3)

write.table(balance1.f[[1]],"NSDUH_balance_unwted_females.csv",sep=",")
write.table(balance1.m[[1]],"NSDUH_balance_unwted_males.csv",sep=",")
write.table(balance1.f[[2]],"NSDUH_balance_wted_females.csv",sep=",")
write.table(balance1.m[[2]],"NSDUH_balance_wted_males.csv",sep=",")

################################################################
#Step 4 - Estimating the treatment effects
################################################################

#Assign PS weights to dataset
#Final weight includes the survey weight times the PS weight
data.female$psw <- get.weights(ps1.f, stop.method="ks.max")
data.male$psw <- get.weights(ps1.m, stop.method="ks.max")

#Combine data
data=rbind(data.female,data.male)

library(margins)
library(survey)
design.ps <- svydesign(ids=~1, weights=~psw, data=data)

outcome.model.psw.logit <- svyglm(cigmon ~
lgb_flag+female+lgb_flag*female+age+race+educ+income+employ, design=design.ps,family =
quasibinomial(link = "logit"))
summary(outcome.model.psw.logit)

#This command uses margins to estimate our TEs within each level of the moderators
summary(margins(outcome.model.psw.logit, at = list(female = c(0, 1)), variables =
"lgb_flag",design=design.ps))

################################################################
#Step 5 - Assessing sensitivity to unobserved confounding
################################################################
```

```r
library(OVtool)

set.seed(24)

#Females

results = OVtool::outcome_model(ps_object = NULL,
              stop.method = NULL,
              data = data.female,
              weights = "psw",
              treatment = "lgb_flag",
              outcome = "cigmon",
              model_covariates = c("age","race","educ","income","employ"),
              estimand = "ATE")

summary(results$mod_results)

ovtool_results_twang = ov_sim(model_results=results,
              plot_covariates=c("age","race","educ","income","employ"),
              es_grid = NULL,
              rho_grid = seq(0, 0.40, by = 0.05),
              n_reps = 100,
              progress = TRUE)
plot.ov(ovtool_results_twang, col='color', print_graphic = '3', p_contours = c(0.05))

#Males

results = OVtool::outcome_model(ps_object = NULL,
              stop.method = NULL,
              data = data.male,
              weights = "psw",
              treatment = "lgb_flag",
              outcome = "cigmon",
              model_covariates = c("age","race","educ","income","employ"),
              estimand = "ATE")

summary(results$mod_results)

ovtool_results_twang = ov_sim(model_results=results,
              plot_covariates=c("age","race","educ","income","employ"),
              es_grid = seq(-0.80, 0.80, by = 0.05),
              rho_grid = seq(0, 0.40, by = 0.05),
              n_reps = 100,
              progress = TRUE)

plot.ov(ovtool_results_twang, col='color', print_graphic = '3', p_contours = c(0.05))
```

**STATA Code**

```
*********************************************************************
***This is the Stata code to run the needed steps in the tutorial
***Uses the NSDUH case study which can be downloaded from our supplementary data files
*********************************************************************

*********************************************************************
*** Step 1 - check for overlap concerns
*********************************************************************

use "data_appendix.dta"

*** Since everything is categorical - we simply need to check for empty cells
*** Create tables stratified by lgb_flag and female

table (age) (lgb_flag female), nototals
table (race) (lgb_flag female), nototals
table (educ) (lgb_flag female), nototals
table (income) (lgb_flag female), nototals
table (employ) (lgb_flag female), nototals

#If there was a continuous covariate, check the minimum and maximum for each level of the moderator
#summary(data.female$continuous.covariate)
#summary(data.male$continuous.covariate)

*********************************************************************
*** Steps 2 & 3 - Estimate PS weights using TWANG within levels of the moderator
*** and assess balance within levels of the moderator
*********************************************************************
*** Must specify adopath where twang STATA ado files have been downloaded
adopath + "/Users/folder/twang STATA/adofiles"

*** subset original data to females only
keep if female == 1
save data_female.dta

use "data_female.dta"
*** Call twang ps function
*** Note that NSDUH sampling weights are controlled for by using the sampw() option

ps lgb_flag i.age i.race i.educ i.income i.employ, ///
  ntrees(10000) stopmethod(ks.max) estimand(ATE) sampw(analwt_c) ///
  rcmd(/usr/local/bin/RScript) ///
  objpath(/Users/folder/project_results) ///
  plotname(/Users/folder/project_results/plot_female.pdf)
*** Examine balance figures (saved as pdf file)
```

*** Examine balance table
balance, unweighted weighted
*** Save twang-generated dataset with propensity score weights
*** Final weight includes the survey weight times the PS weight
save data_female_wgts

*** Subset original data to males only
use "data_appendix.dta"
keep if female == 0
save data_male.dta

use "data_male.dta"
*** Call twang ps function
ps lgb_flag i.age i.race i.educ i.income i.employ, ///
  ntrees(10000) stopmethod(ks.max) estimand(ATE) sampw(analwt_c) ///
  rcmd(/usr/local/bin/RScript) ///
  objpath(/Users/folder/project_results) ///
  plotname(/Users/folder/project_results/plot_male.pdf)
*** Examine balance figures (saved as pdf file)
*** Examine balance table
balance, unweighted weighted
*** Save twang-generated dataset with propensity score weights
*** Final weight includes the survey weight times the PS weight

save data_male_wgts

*******************************************************************
*** Step 4 - Estimating the treatment effects
*******************************************************************

*** Combine data_male_wgts.dta and data_female_wgts.dta -- will run outcome model
*** on combined dataset

append using "data_female_wgts.dta"
save "data_combined_wgts.dta"

*** Outcome model: Propensity score-weighted logistic regression
svyset [pweight=ksmaxate]
svy: logit cigmon lgb_flag#female i.age i.race i.educ i.income i.employ
margins, dydx(lgb_flag) over(female)

*******************************************************************
*******************************************************************
*******************************************************************

Appendix B: Code to run for a generic data set

**R Code**

```
#This is the R code to run the needed steps in the tutorial
#Assumes a generic data structure with the following key variables:
#outcome, treatment, moderator, categorical and continuous covariates

library(twang)

#Subset data down to the different levels of the moderator

data1=data[data$moderator==1,]
data0=data[data$moderator==0,]

#########################################
#Step 1 - check for overlap concerns
#########################################

#For categorical confounders check for empty cells
table(data1$treatment,data1$categorical.confounder1)
table(data1$treatment,data1$categorical.confounder2)
table(data1$treatment,data1$categorical.confounder3)

table(data0$treatment,data0$ categorical.confounder1)
table(data0$treatment,data0$ categorical.confounder2)
table(data0$treatment,data0$ categorical.confounder3)

#For continuous covariate, check the minimum and maximum for each level of the moderator
summary(data1$continuous.covariate1)
summary(data0$continuous.covariate1)

summary(data1$continuous.covariate2)
summary(data0$continuous.covariate2)

summary(data1$continuous.covariate3)
summary(data0$continuous.covariate3)

##############################################################################
#Step 2 - estimate PS weights using TWANG within levels of the moderator
##############################################################################

set.seed(175659)

#Note that sampling weights can be controlled for by using the sampw option

ps.1 <- ps(treatment ~categorical.confounder1+categorical.confounder2+categorical.confounder3+
        continuous.confounder1+continuous.confounder2+continuous.confounder3, data = data1,
        n.trees=10000,interaction.depth=2,shrinkage=0.01,perm.test.iters=0,
        stop.method=c("ks.max"),estimand = "ATE",verbose=FALSE,sampw = NA)

ps.0<- ps(treatment ~ categorical.confounder1+categorical.confounder2+categorical.confounder3+
        continuous.confounder1+continuous.confounder2+continuous.confounder3, data = data0,
```

```r
                n.trees=10000,interaction.depth=2,shrinkage=0.01,perm.test.iters=0,
                stop.method=c("ks.max"),estimand = "ATE",verbose=FALSE,sampw = NA )
```

######################################################################
#Step 3 - assess balance within levels of the moderator
######################################################################

```r
plot(ps.1)
balance1 <- bal.table(ps.1)
balance1
summary(ps.1)

plot(ps.0)
balance0 <- bal.table(ps.0)
balance0
summary(ps.0)

plot(ps.1,plots=1)
plot(ps.0,plots=1)

plot(ps.1,plots=2)
plot(ps.0,plots=2)

plot(ps.1,plots=3)
plot(ps.0,plots=3)

write.table(balance1[[1]],"balance_unwted_moderator1.csv",sep=",")
write.table(balance0[[1]],"balance_unwted_moderator0.csv",sep=",")
write.table(balance1[[2]],"balance_wted_moderator1.csv",sep=",")
write.table(balance0[[2]],"balance_wted_moderator0.csv",sep=",")
```

######################################################################
#Step 4 - Estimating the treatment effects
######################################################################

```r
#Assign PS weights to dataset
#Final weight includes the survey weight times the PS weight
data1$psw <- get.weights(ps.1, stop.method="ks.max")
data0$psw <- get.weights(ps.0, stop.method="ks.max")

#Combine data
data=rbind(data1,data0)

library(margins)
library(survey)
design.ps <- svydesign(ids=~1, weights=~psw, data=data)

outcome.model.psw.logit <- svyglm(outcome~treatment+moderator+treatment*moderator+
categorical.confounder1+categorical.confounder2+categorical.confounder3+continuous.confounder1+
continuous.confounder2+continuous.confounder3, design=design.ps,family =  quasibinomial(link =
"logit"))
summary(outcome.model.psw.logit)

#This command uses margins to estimate our TEs within each level of the moderators
summary(margins(outcome.model.psw.logit, at = list(moderator = c(0, 1)), variables =
"treatment",design=design.ps))
```

```
########################################################################
#Step 5 - Assessing sensitivity to unobserved confounding
########################################################################

library(OVtool)

set.seed(24)

#Moderator Level 1

results = OVtool::outcome_model(ps_object = NULL,
                    stop.method = NULL,
                    data = data1,
                    weights = "psw",
                    treatment = "treatment",
                    outcome = "outcome",
                    model_covariates =c("categorical.confounder1","categorical.confounder2",
                        "categorical.confounder3","continuous.confounder1","continuous.confounder2",
                        "continuous.confounder3"),
                    estimand = "ATE")

summary(results$mod_results)

ovtool_results_twang = ov_sim(model_results=results,
                    plot_covariates =c("categorical.confounder1","categorical.confounder2",
                        "categorical.confounder3","continuous.confounder1","continuous.confounder2",
                        "continuous.confounder3"),
                    es_grid = NULL,
                    rho_grid = seq(0, 0.40, by = 0.05),
                    n_reps = 100,
                    progress = TRUE)

plot.ov(ovtool_results_twang, col='color', print_graphic = '3', p_contours = c(0.05))

#Moderator Level 0

results = OVtool::outcome_model(ps_object = NULL,
                    stop.method = NULL,
                    data = data0,
                    weights = "psw",
                    treatment = "treatment",
                    outcome = "outcome",
                    model_covariates =c("categorical.confounder1","categorical.confounder2",
                        "categorical.confounder3","continuous.confounder1","continuous.confounder2",
                        "continuous.confounder3"),
                    estimand = "ATE")

summary(results$mod_results)

ovtool_results_twang = ov_sim(model_results=results,
                    plot_covariates =c("categorical.confounder1","categorical.confounder2",
                        "categorical.confounder3","continuous.confounder1","continuous.confounder2",
                        "continuous.confounder3"),
```

```
                es_grid = seq(-0.80, 0.80, by = 0.05),
                rho_grid = seq(0, 0.40, by = 0.05),
                n_reps = 100,
                progress = TRUE)

plot.ov(ovtool_results_twang, col='color', print_graphic = '3', p_contours = c(0.05))
```

**STATA Code**

```
**********************************************************************
***This is the Stata code to run the needed steps in the tutorial
***Assumes a generic data structure with the following key variables:
***outcome, treatment, moderator, categorical and continuous covariates
**********************************************************************

**********************************************************************
*** Step 1 - check for overlap concerns
**********************************************************************

use "data.dta"

*** For categorical confounders check for empty cells
*** Create tables stratified by treatment and moderator

table (categorical.confounder1) (treatment moderator), nototals
table (categorical.confounder2) (treatment moderator), nototals
table (categorical.confounder3) (treatment moderator), nototals

#For continuous covariate, check the minimum and maximum for each level of the moderator
summary(data1$continuous.covariate1)
summary(data0$continuous.covariate1)

summary(data1$continuous.covariate2)
summary(data0$continuous.covariate2)

summary(data1$continuous.covariate3)
summary(data0$continuous.covariate3)

**********************************************************************
*** Steps 2 & 3 - Estimate PS weights using TWANG within levels of the moderator
*** and assess balance within levels of the moderator
**********************************************************************
*** Must specify adopath where twang STATA ado files have been downloaded
adopath + "/Users/folder/twang STATA/adofiles"

*** subset original data to moderator level 1 only
keep if moderator == 1
save data_1.dta

use "data_1.dta"
*** Call twang ps function
*** Note that sampling weights can be controlled for by using the sampw option
```

```
ps treatment i.categorical.confounder1 i.categorical.confounder2 i.categorical.confounder3
  continuous.confounder1 continuous.confounder2 continuous.confounder3, ///
  ntrees(10000) stopmethod(ks.max) estimand(ATE) sampw(NA) ///
  rcmd(/usr/local/bin/RScript) ///
  objpath(/Users/folder/project_results) ///
  plotname(/Users/folder/project_results/plot_moderator1.pdf)
*** Examine balance figures (saved as pdf file)
*** Examine balance table
balance, unweighted weighted
*** Save twang-generated dataset with propensity score weights
*** Final weight includes the survey weight times the PS weight
save data_1_wgts

*** Subset original data to those with moderator = 0 only
use "data_appendix.dta"
keep if moderator == 0
save data_0.dtadata_0.dta

use "data_0.dta"
*** Call twang ps function
ps treatment i.categorical.confounder1 i.categorical.confounder2 i.categorical.confounder3
  continuous.confounder1 continuous.confounder2 continuous.confounder3, ///
  ntrees(10000) stopmethod(ks.max) estimand(ATE) sampw(NA) ///
  rcmd(/usr/local/bin/RScript) ///
  objpath(/Users/folder/project_results) ///
  plotname(/Users/folder/project_results/plot_0.pdf)
*** Examine balance figures (saved as pdf file)
*** Examine balance table
balance, unweighted weighted
*** Save twang-generated dataset with propensity score weights
*** Final weight includes the survey weight times the PS weight

save data_0_wgts

********************************************************************
*** Step 4 - Estimating the treatment effects
********************************************************************

*** Combine data_0_wgts.dta and data_1_wgts.dta -- will run outcome model
*** on combined dataset

append using "data_1_wgts.dta"
save "data_combined_wgts.dta"

*** Outcome model: Propensity score-weighted logistic regression
svyset [pweight=ksmaxate]
svy: logit outcome treatment#moderator i.categorical.confounder1 i.categorical.confounder2
i.categorical.confounder3 continuous.confounder1 continuous.confounder2 continuous.confounder3
margins, dydx(treatment) over(moderator)

********************************************************************
********************************************************************
********************************************************************
```